\documentclass{SciPost}

\hypersetup{
    colorlinks,
    linkcolor={red!50!black},
    citecolor={blue!50!black},
    urlcolor={blue!80!black}
}

\usepackage[bitstream-charter]{mathdesign}
\usepackage{subcaption}
\DeclareSymbolFont{usualmathcal}{OMS}{cmsy}{m}{n}
\DeclareSymbolFontAlphabet{\mathcal}{usualmathcal}

\fancypagestyle{SPstyle}{
\fancyhf{}
\lhead{\colorbox{scipostdeepblue}{\bf \color{white} ~SciPost Physics Proceedings }}
\rhead{{\bf \color{scipostdeepblue} ~Submission }}

\fancyfoot[C]{\textbf{\thepage}}
}

\begin{document}

\pagestyle{SPstyle}

\begin{center}{\Large \textbf{\color{scipostdeepblue}{
Real-Time Motion Correction in Magnetic Resonance Spectroscopy: AI solution inspired by fundamental science\\
}}}\end{center}

\begin{center}\textbf{
Alberto Annovi\textsuperscript{1}, 
\underline{Benedetta Argiento}\textsuperscript{1,2$\star$}, 
Silvia Capuani\textsuperscript{3}, 
Matteo Cacioppo\textsuperscript{4}, 
Andrea Ciardiello\textsuperscript{1,5}, 
Roberto Coccurello\textsuperscript{3}, 
Stefano Giagu\textsuperscript{1,4}, 
Federico Giove\textsuperscript{4,5}, 
Alessandro Lonardo\textsuperscript{1}, 
Francesca Lo Cicero\textsuperscript{1}, 
Alessandra Maiuro\textsuperscript{1,4}, 
Carlo Mancini Terracciano\textsuperscript{1}, 
Mario Merola\textsuperscript{1,2}, 
Marco Montuori\textsuperscript{3}, 
Emilia Nisticò\textsuperscript{4}, 
Pierpaolo Perticaroli\textsuperscript{1}, 
Biagio Rossi\textsuperscript{1}, 
Cristian Rossi\textsuperscript{1}, 
Elvira Rossi\textsuperscript{1,2}, 
Francesco Simula\textsuperscript{4}, 
and Cecilia Voena\textsuperscript{1}
}\end{center}

\begin{center}

{\bf 1} Istituto di Fisica Nucleare, Italia  
\\
{\bf 2} Università degli Studi di Napoli Federico II, Napoli, Italia  
\\
{\bf 3} Consiglio Nazionale delle Ricerche, ISC, Roma, Italia 
\\
{\bf 4} Sapienza Università di Roma, Roma, Italia  
\\
{\bf 5} Istituto Superiore di Sanità, Roma, Italia 
\\
{\bf 4} Centro Ricerche Enrico Fermi, Roma, Italia 
\\
{\bf 5} Fondazione Santa Lucia IRCCS, Roma, Italia

\vspace{\baselineskip}
$\star$ \href{mailto:email1}{\small bargient@cern.ch}\quad
\end{center}

\definecolor{palegray}{gray}{0.95}
\begin{center}
\colorbox{palegray}{
  \begin{tabular}{rr}
  \begin{minipage}{0.37\textwidth}
    \includegraphics[width=60mm]{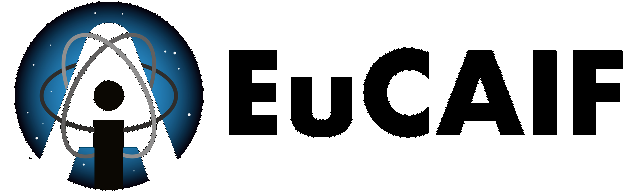}
  \end{minipage}
  &
  \begin{minipage}{0.5\textwidth}
    \vspace{5pt}
    \vspace{0.5\baselineskip} 
    \begin{center} \hspace{5pt}
    {\it The 2nd European AI for Fundamental \\Physics Conference (EuCAIFCon2025)} \\
    {\it Cagliari, Sardinia, 16-20 June 2025
    }
    \vspace{0.5\baselineskip} 
    \vspace{5pt}
    \end{center}
    
  \end{minipage}
\end{tabular}
}
\end{center}

\section*{\color{scipostdeepblue}{Abstract}}

\textbf{Magnetic Resonance Spectroscopy (MRS) is a powerful non-invasive tool for metabolic tissue analysis but is often degraded by patient motion, limiting clinical utility. The RECENTRE project (REal-time motion CorrEctioN in magneTic Resonance) presents an AI-driven, real-time motion correction pipeline based on optimized GRU networks, inspired by tagging and fast-trigger algorithms from high-energy physics. Models evaluated on held-out test sets achieve good predictive performance (R$^2$>0.87) and overall positive framewise displacement (FD) gains. These results demonstrate feasibility for prospective scanner integration; future work will complete in-vivo validation.}


\vspace{\baselineskip}

\noindent\textcolor{white!90!black}{%
\fbox{\parbox{0.975\linewidth}{%
\textcolor{white!40!black}{\begin{tabular}{lr}%
  \begin{minipage}{0.6\textwidth}%
    {\small Copyright attribution to authors. \newline
    This work is a submission to SciPost Phys. Proc. \newline
    License information to appear upon publication. \newline
    Publication information to appear upon publication.}
  \end{minipage} & \begin{minipage}{0.4\textwidth}
    {\small Received Date \newline Accepted Date \newline Published Date}%
  \end{minipage}
\end{tabular}}
}}
}


\vspace{10pt}
\noindent\rule{\textwidth}{1pt}
\tableofcontents
\noindent\rule{\textwidth}{1pt}
\vspace{10pt}


\section{Introduction}
\label{sec:intro}

Magnetic Resonance (MR) is a well-established, non-invasive modality for the study of tissue structure and function. Magnetic Resonance Spectroscopy (MRS)\cite{Saleh2020_MotionCorrectionMRS} extends MR to metabolic and biochemical assessment, with potential clinical applications. However, patient motion remains a critical obstacle: even modest displacements produce spectral distortions, baseline shifts and loss of quantification that reduce reproducibility and limit diagnostic value.

The RECENTRE project proposes a real-time motion correction \cite{Andronesi2020_MotionCorrectionMRS} \cite{Keating2010_ProspectiveMRS} \cite{Marsman2021_ProspectiveCorrectionMRS} pathway for MRS based on compact deep recurrent networks. The design is inspired by algorithmic techniques from high-energy physics, such as tagging and fast-triggering \cite{Giagu2020_FastGRN}, which enable low-latency decisions on streaming data. 
Training and evaluation emphasise two complementary goals. First, the model must accurately \cite{Tamada2020_NoiseDL} predict motion-related parameters that can be used for prospective adjustment during the acquisition. Second, the training objective explicitly favours reductions in framewise displacement while preserving spectral fidelity.

\section{Neural Network and Framewise Displacement approach}
\label{sec:methods}
Recurrent neural networks (RNNs) are well suited to handle sequential data, as they capture temporal dependencies across consecutive observations. Among their variants, Gated Recurrent Units (GRUs) provide comparable performance to long short-term memory (LSTM) networks while requiring fewer parameters and less training time. This balance reduces overfitting risks and improves efficiency, making GRUs an appropriate choice for real-time applications. Within the RECENTRE project, a GRU network was adopted to predict motion corrections directly from short temporal sequences of MR acquisitions.  

\subsection*{Dataset}
\label{dataset}
\begin{figure}[h!]
    \centering
    \includegraphics[width=0.65\linewidth]{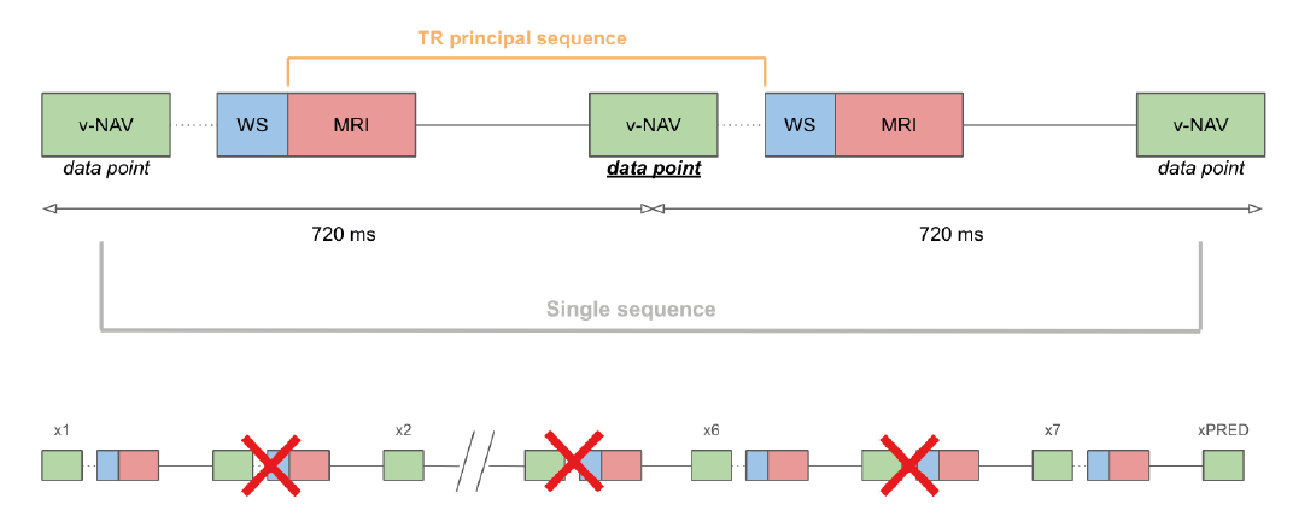}
    \caption{Each input sample consists of 7 sequences, each covering two consecutive acquisition points (for a total of 14 time points, given the short TR). The model is trained to predict the subsequent 15th point.}
    \label{fig:input_seq}
\end{figure}

The data used for training were obtained from the Human Connectome Project (HCP)\cite{HCP_S1200_ReferenceManual}. A total of 1113 subjects were scanned on Siemens 3T MRI systems with a repetition time (TR) of 720 ms and an echo time (TE) of 33.1 ms.  Three acquisition types have been included in this work, differing in the number of frames per sequence: Resting State (1200 frames), Working Memory (316 frames), and Language (405 frames). The motion parameters used as input to the network consist of three translations and three rotations, extracted through rigid realignment during post-processing of the MRI data.  As illustrated in Fig.~\ref{fig:input_seq}, the network input was formed by 7 sequences of two data points each (14 in total), then the model predicts the following 15th point.

\subsection*{Framewise displacement gain}
Framewise displacement (FD) was employed as a subject-specific index of motion, providing a scalar measure of head movement at each time point. FD was computed both from the ground-truth and the predicted motion parameters. The FD gain was then defined as:

\begin{equation}
FD_{total} = \sum |T_{i,t} - T_{i,t+1}| + 50 \cdot \frac{\pi}{180} \sum |R_{i,t} - R_{i,t+1}|
\end{equation}

\begin{equation}
FD_{predicted} = \sum |\hat{T}_{i,t+1} - T_{i,t+1}| + 50 \cdot \frac{\pi}{180} \sum |\hat{R}_{i,t+1} - R_{i,t+1}|
\end{equation}

\begin{equation}
FD_{gain} = \frac{FD_{total} - FD_{predicted}}{FD_{total}}
\label{eq:fd_gain}
\end{equation}

where $T$ are the translation parameters (mm), $R$ the rotation parameters (rad), and $\hat{T}, \hat{R}$ the predicted quantities. A positive $FD_{gain}$ indicates an improvement, i.e. reduced motion relative to the original sequence.

\subsection*{Neural network model and training objective}
The adopted model is composed of a GRU layer with hidden size 128, followed by normalization, non-linear activations and two fully connected layers. The total number of trainable parameters is approximately 270k. The network outputs the predicted motion parameters together with their associated uncertainty estimates.  

\begin{figure}[h]
    \centering
    \begin{subfigure}[b]{0.32\textwidth}
        \centering
        \includegraphics[width=\linewidth]{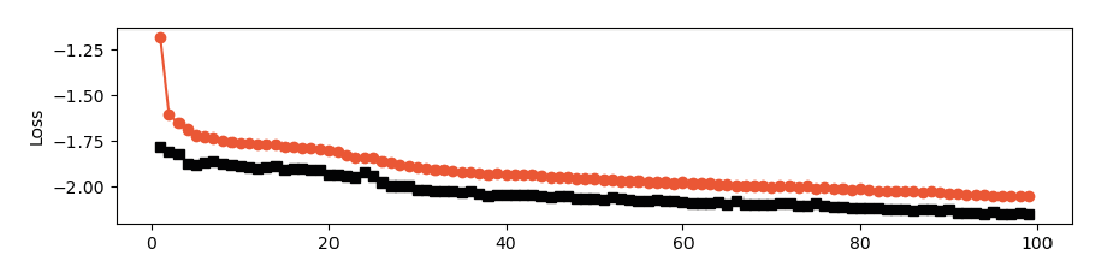}
        \caption{Resting State}
        \label{fig:training_rs}
    \end{subfigure}
    \hfill
    \begin{subfigure}[b]{0.32\textwidth}
        \centering
        \includegraphics[width=\linewidth]{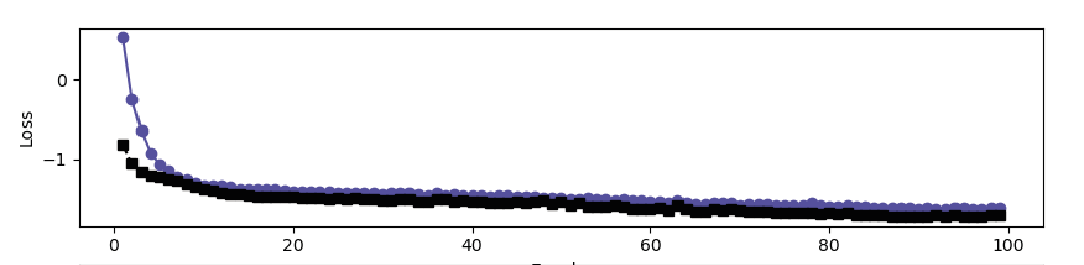}
        \caption{Working Memory}
        \label{fig:training_wm}
    \end{subfigure}
    \hfill
    \begin{subfigure}[b]{0.32\textwidth}
        \centering
        \includegraphics[width=\linewidth]{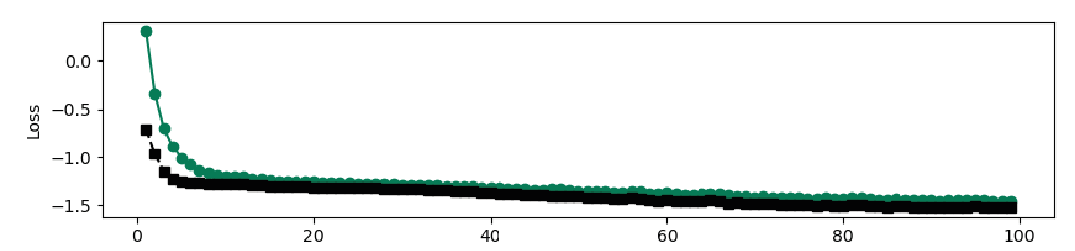}
        \caption{Language}
        \label{fig:training_lang}
    \end{subfigure}
    \caption{Training and validation loss curves for the three acquisition types.}
    \label{fig:loss_curves}
\end{figure}

Training was performed with the Adam optimizer and early stopping based on validation performance. The loss function combined a probabilistic likelihood term with an explicit penalty on motion reduction:

\begin{equation}
\mathcal{L} = \text{NLL} - 0.1 \times FD_{gain}
\end{equation}

where NLL denotes the negative log-likelihood of the predicted parameters, and the FD term penalizes negative FD gains. This design enforces both predictive accuracy and effective reduction of head motion. The corresponding training and validation loss curves for the three acquisition types are reported in Fig.~\ref{fig:loss_curves}.

\section{Results}
\label{sec:results}

Results were obtained on the held-out test set across the three acquisition types 
(Resting State, Working Memory, and Language), as described in Section~\ref{sec:methods}.

Training curves demonstrate stable convergence of the loss across all acquisition types. 
No significant overfitting was observed. 

\begin{figure}[h]
    \centering
    \includegraphics[width=0.99\linewidth]{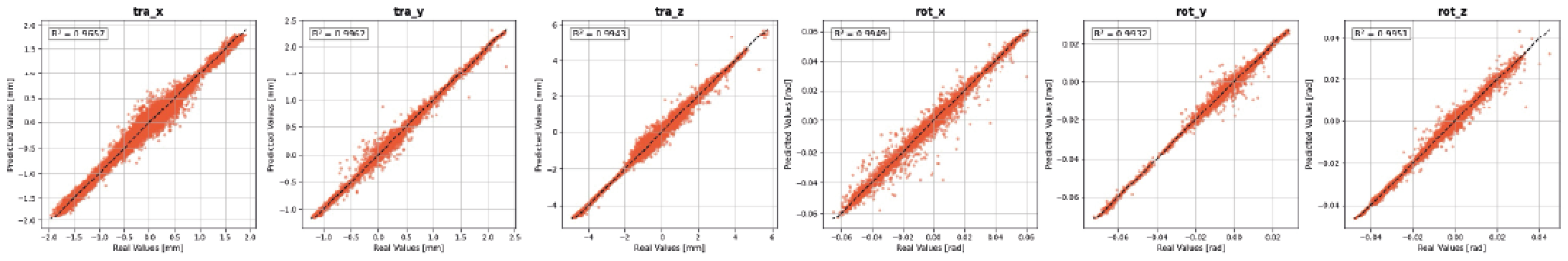}
    \caption{Predicted vs.\ true motion parameters on the test set for the Resting State acquisition. The figure reports the six estimated motion parameters, namely the three rotational displacements (expressed in radians) and the three rotational angles (expressed in millimeters).}
    \label{fig:pred_vs_true}
\end{figure}

Prediction accuracy was evaluated by comparing predicted motion parameters against the ground truth. 
Scatter plots of predicted vs.\ true displacements showed good alignment for both translations and rotations, as seen in Fig. \ref{fig:pred_vs_true}. 
The coefficient of determination ($R^2$) consistently exceeded 0.87 across tasks, 
confirming the reliability of the network predictions.

\begin{figure}[h]
    \centering
    \begin{subfigure}[b]{0.32\textwidth}
        \centering
        \includegraphics[width=\linewidth]{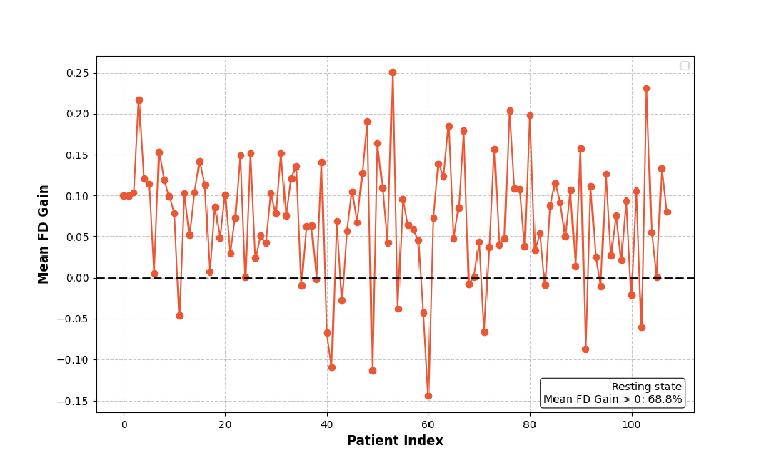}
        \caption{Resting State}
        \label{fig:fdgain_rs}
    \end{subfigure}
    \hfill
    \begin{subfigure}[b]{0.32\textwidth}
        \centering
        \includegraphics[width=\linewidth]{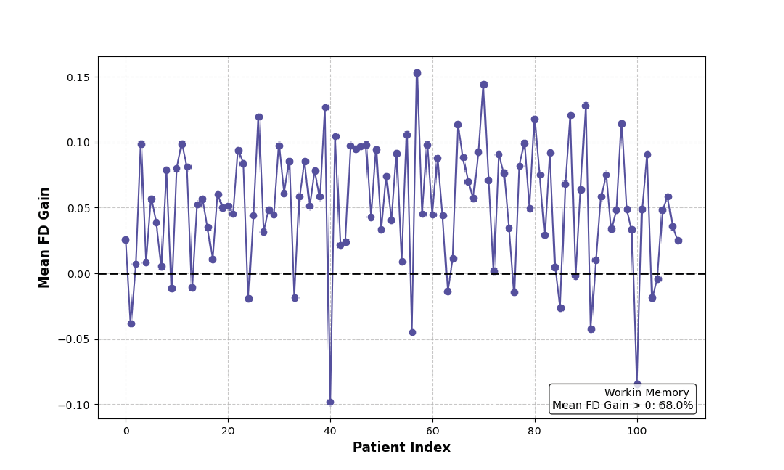}
        \caption{Working Memory}
        \label{fig:fdgain_wm}
    \end{subfigure}
    \hfill
    \begin{subfigure}[b]{0.32\textwidth}
        \centering
        \includegraphics[width=\linewidth]{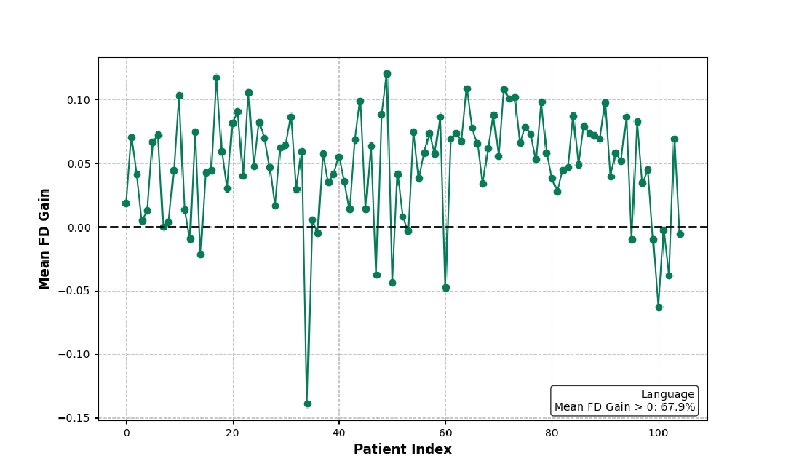}
        \caption{Language}
        \label{fig:fdgain_lang}
    \end{subfigure}
    \caption{Mean FD gain per patient (average over all predictions). Negative FD gains indicate that the network sometimes slightly overestimates motion.}
    \label{fig:fd_gain}
\end{figure}

The effectiveness of the network in reducing apparent motion was quantified using the FD gain metric 
(see Eq.~\eqref{eq:fd_gain}). Positive FD gains were consistently observed across all acquisition  (Fig. \ref{fig:fd_gain}, 
demonstrating improved motion estimates. Results were comparable among Resting State, Working Memory, and Language tasks.

\section{Conclusions}

The GRU-based predictor achieves consistently high goodness-of-fit ($R^2>0.87$) across evaluated dimensions, capturing both large and subtle motion patterns. FD-gain analysis shows mostly positive values, reflecting a net reduction of estimated motion for the majority of patients, with only occasional slight overestimation. The GRU network will be integrated into the Siemens syngo MR MAGNETOM workflow, where the Siemens Image Calculation Environment (ICE) will use the predicted roto-translation parameters for motion-corrected image reconstruction. The model will run on the Siemens Framework for Image Reconstruction Environment (FIRE) with deployment on the Siemens MARS workstation, leveraging on-board GPUs to accelerate prediction. Validation will be performed on in-vivo data acquired with the Siemens 3T Prisma scanner at IRCCS Santa Lucia in Rome.

\section*{Acknowledgements}
The RECENTRE Project (Prot. P202294JHK) was supported by the PRIN PNRR 2022 program which enabled this research and its presentation.

\bibliography{Proceedings-EuCAIF.bib}









\end{document}